\documentclass[a4paper,10pt]{article}
\usepackage{graphicx}
\usepackage{t1enc}

\begin{document}

\title{Variable stars in the field of open cluster NGC~6939}%
\author{G. Maciejewski\\Centrum Astronomii UMK, ul. Gagarina 11, 87-100 Toru\'n\\gm@astri.uni.torun.pl \\ Ts. Georgiev \\ Institute of Astronomy, Bulgarian Academy of Sciences, \\ 72 Tsarigradsko Chausse Blvd. 1784 Sofia, Bulgaria \\ A. Niedzielski \\ Centrum Astronomii UMK, ul. Gagarina 11, 87-100 Toru\'n}%
\maketitle

\textit{Abstract}: The results of  CCD photometric survey performed with the 90/180 cm Schmidt-Cassegrain Telescope of the Nicolaus Copernicus University Astronomical Observatory   in Piwnice (Poland) and the 70/172 cm Schmidt Telescope  of the National Astronomical Observatory (NAO) at Rozhen (Bulgaria) of the field of 1 Gyr old open cluster NGC~6939 are presented. Twenty two variable stars were detected,  four of them  previously known. Four eclipsing systems (3 detached and 1 contact binary)  were found to be members of the cluster.  Analysis of the brightness of the contact binary V20 strongly supports the distance to the cluster of $1.74 \pm 0.20$ kpc. The small population of contact binaries in NGC 6939 confirms also the relatively young age of the cluster.
 
\textit{Keywords}: Open clusters and associations: individual: NGC 6939 -- stars: variables: general

\section{Introduction}

NGC~6939 is a relatively rich open cluster located in the constellation of Cepheus ($\alpha=20^{\mathrm{h}}31^{\mathrm{m}}30^{\mathrm{s}}$, $\delta=+60^{\circ}39'22''$), $12.3^{\circ}$ above the Galactic plane. The cluster was  subject of numerous investigations. The first publication on NGC~6939 presenting a photographic catalog of 370 cluster's stars is dated on 1923 (Kustner 1923). Cuffey (1944) estimated  distance to the cluster for 1300 pc. Cannon \& Lloyd (1969) obtained the age of about $10^{9}$ yr. However, the first deep CCD photometry was obtained just at the beginning of current century. Rosvick \& Balam (2002) collected BVI photometry for stars located in the central part of the cluster. These authors obtained the age of $(1.6\pm0.3)\cdot10^{9}$ yr and found premises of variability of interstellar extinction within the cluster. The reddening  was found to range from $E(B-V)$=0.29 mag to $E(B-V)$=0.41 mag. In the following study Andreuzzi et al. (2004) obtained the  age of $(1.0-1.3)\cdot10^{9}$ yr, variable reddening $E(B-V)=0.34-0.38$ mag, and distance of 1.8 kpc based on UBVI CCD photometry. More recently, Maciejewski \& Niedzielski (2007), who based on wide-field BV data, obtained the age of 1.25 Gyr, $E(B-V)=0.38^{+0.18}_{-0.10}$ mag, distance modulus $(m-M)=12.15_{-0.72}^{+0.56}$ mag, and distance of $1.56^{+0.64}_{-0.59}$ kpc. Authors also determined cluster's core radius $r_{\mathrm{core}}=2.2'$ and limiting radius $r_{\mathrm{lim}}=15.2'$. The cluster occurred to be larger than previous estimations of the apparent diameter had indicated. Moreover, the evidence of the mass segregation within the cluster was found based on mass function investigation.

The metallicity of NGC~6939 was studied in a few papers (Canterna at al. 1986, Geisler et al. 1991, Thogersen et al. 1993, Fiel et al. 2002) and occurred to be below the solar value with [Fe/H] between $-0.19$ and $-0.10$ depending on authors. The cluster was also searched for variable stars. During 16 nights Robb \& Cardinal (1998) monitored 215 stars located in the central part of the cluster. As a result variability of 6 stars was detected. Three of them were located in the red clump and revealed long term light-curve changes. Remaining next three variables were classified as eclipsing binary systems. 

As the apparent diameter of the cluster was found in Maciejewski \& Niedzielski (2007) much larger than the value previously accepted of 5' (Dias et al. 2002) we performed a dedicated CCD search for variable stars in this cluster. In this paper we present  results of this  survey. The main goal of this study was to constrain the open cluster parameters better through investigation of  the population of variable stars, members of  the cluster.

\section{Observation and data reduction}

The field of NGC~6939 was searched for variable stars during two campaigns performed in BV filters with two Schmidt telescopes. The first campaign was performed between   March 22--24 and May 2--4, 2006 with the 90/180 cm Schmidt-Cassegrain Telescope of the Nicolaus Copernicus University Astronomical Observatory   in Piwnice near Toru\'n, Poland. The telescope was used in the imaging mode with a 60 cm correction plate  and a field-flattening lens mounted near the focal plane.  SBIG STL-11000 CCD camera (4008 $\times$ 2672 pixels $\times$ 9 $\mu$m) was used as a detector. The field of view was 72 arcmin in declination and 48 arcmin in right ascension with the scale of 1.08 arcsec per pixel. The $2 \times 2$ binning was used to increase the signal-to-noise ratio. The exposure time  was set to 600 s and typical seeing (FWHM) was 5--6''. During about 23 hours  of observations in total 98 images in V and 30 in B were obtained. Over 6950 stars between 13.5 and 18.5 mag in V were monitored.

The second campaign was performed during 7 nights in 2007, March 14 -- 18, May 11, 13--15, and June 11 with the 70/172 cm Schmidt Telescope  of the National Astronomical Observatory (NAO) at Rozhen (Bulgaria), operated by the Institute of Astronomy of the Bulgarian Academy of Sciences. The instrument was equipped with a 50 cm correction plate  and an SBIG ST-8 CCD camera (1530 $\times$ 1020 pixels $\times$ 9 $\mu$m). The field of view was 27 arcmin in declination and 18 arcmin in right ascension during March observations and 18 arcmin in declination and 27 arcmin in right ascension during remaining nights. The scale was 1.08 arcsec per pixel. The  exposure time was 300 s with typical seeing (FWHM) of 3''. During almost 17 hours of monitoring 99 images in V and 7 in B were acquired. Due to smaller field of view only 2000 stars as faint as $V=19.5$ mag were observed.

The collected observations were reduced with the software pipeline developed for the Semi-Automatic Variability Search\footnote{http://www.astri.uni.torun.pl/\~{}gm/SAVS} sky survey (Niedzielski et al. 2003, Maciejewski \& Niedzielski 2005). CCD frames were processed with a standard procedure including subtraction of dark frames and dividing by sky-flat-field frames. Instrumental stellar magnitudes were derived by the means of differential aperture photometry against selected standard star which magnitudes were taken from photometric database available at the Open Cluster Survey web site\footnote{http://www.astri.uni.torun.pl/\~{}gm/OCS} (OCS, Maciejewski \& Niedzielski 2007). The aperture diameter was calculated for individual objects as $3\sigma$ of the estimated Gaussian stellar profile. The instrumental coordinates of stars were transformed into equatorial ones based on positions of stars brighter than 16 mag extracted from the Guide Star Catalog. The candidates for new variable stars were selected from V-band database using the analysis of variance method (ANOVA, Schwarzenberg-Czerny 1996). The accuracy of our photometry was 0.025 mag for stars brighter than 16.0 mag in V and gradually decreased reaching 0.15 mag for the faintest stars of $V=18.25$ mag and $V=19.25$ mag for observations from Piwnice and Rozhen, respectively. The search for candidates was performed on all detected stars what allowed us to find variables with very small amplitudes, comparable to our photometric precision. All variables were verified after careful visual inspection. 

\section{Results}

\begin{table*}
\centering
\caption{List of variable stars detected in the field of NGC~6939. $V_{\mathrm{max}}$ -- maximal brightness in V band, $\Delta V$ -- amplitude of variation in V, $(B-V)_{\mathrm{max}}$ -- color index at the maximum of brightness, $P$ -- period of variation, $T_{0}$ -- epoch of minimum brightness for eclipsing systems or maximum for pulsating stars, $d$ -- distance from the cluster center, types of variability: EA -- eclipsing binary of Algol type, EB -- semidetached binary, EW -- contact system, DSCT -- pulsating star of $\delta$Scuti type, RRAB -- pulsating star of RR Lyrae type with asymmetric light curve, DCEP -- pulsating star of $\delta$Cephei type, MISC --  miscellaneous variable of unresolved type.}
\label{table1}
\begin{tabular}{llccccccll}
\hline
ID & Coordinates   & $V_{\mathrm{max}}$ & $\Delta V$ & $(B-V)_{\mathrm{max}}$ & $P$ & $T_{0}$ & $d$ & Type & membership\\
   & J2000.0       & (mag)              & (mag)      & (mag)                  & (days) & HJD--2450000 & (arcmin) &   &\\
\hline
V1 & 202823+611009 & 15.36 & 0.41 & 1.09 & 0.4315 & 3817.8438 & 38.4 & EW & no\\
V2 & 202828+601107 & 15.00 & 0.13 & 0.65 & 0.0470 & 3817.6311 & 36.0 & DSCT& no\\
V3 & 202927+601405 & 15.25 & 0.25 & 0.91 & 0.4374 & 3817.6651 & 29.4 & EW&  no\\
V4 & 202932+601139 & 17.94 & 0.90 & 1.21 & 0.2671 & 3818.3561 & 31.3 & EW& no\\
V5 & 202935+603833 & 16.03 & 0.86 & 0.87 & 0.5375 & 3818.4959 & 14.1 & RRAB& unlikely\\
V6 & 203001+604602 & 16.35 & 0.30 & 1.25 & 0.3599 & 3817.9750 & 12.8 & EW& unlikely\\
V7 & 203004+603432 & 18.44 & 0.68 & 1.19 & 0.2567 & 4232.9223 & 11.6 & EW& unlikely\\
V8 & 203016+603631 & 17.73 & 0.44 & 1.56 & 0.2419 & 3817.5655 & 9.5 & EW& unlikely\\
V9 & 203104+601615 & 17.37 & 0.54 & 1.09 & 0.3141 & 3817.7906 & 23.3 & EW& no\\
V10 & 203104+604324 & 16.24 & 0.13 & 1.24 & 0.8940 & 3819.2161 & 5.1 & RR:& unlikely\\
V11 & 203113+603647 & 16.14 & 0.24 & 1.20 & 0.3549 & 3817.8614 & 3.3 & EB& unlikely\\
V12 & 203118+603809 & 14.33 & 0.22 & 0.89 & 4.9590: & 3827.1868 & 1.9 & EA& likely\\
V13 & 203124+604017 & 13.54 & 0.11 & 0.79 & 1.6144 & 4179.5658 & 1.2 & DCEP& unlikely\\
V14	& 203129+603934	& 14.88 & 0.11 & 0.89 & 3.5974 & 4183.4491 & 0.2 & EA & likely\\
V15	& 203140+602836	& 14.21 & 0.52 & 1.88	& -- & -- & 10.8 & MISC & no\\
V16 & 203203+600632 & 15.85 & 0.35 & 0.76 & 0.0896 & 3817.7129 & 33.1 & DSCT& no\\
V17 & 203214+604157 &	16.23 & 0.30 & 1.17 & 1.1879 & 3817.7965 & 6.0 & EA& likely\\	
V18 & 203245+603555 & 15.21 & 0.44 & 2.09 & -- & -- & 9.8 & MISC& no\\
V19 & 203308+602006 & 13.95 & 0.54 & 1.01 & 0.3590 & 3817.8847 & 22.7 & EW& no\\
V20 & 203322+603714 & 17.02 & 0.56 & 1.10 & 0.2951 & 3817.7000 & 13.9 & EW& yes\\
V21 & 203326+610348 & 16.93 & 0.48 & 0.83 & 0.7014 & 3818.2916 & 28.3 & EB& no\\
V22 & 203442+603244 & 16.59 & 0.50 & 0.85 & 0.3223 & 3818.0826 & 24.4 & RRAB& no\\
\hline
\end{tabular}
\end{table*}

\begin{figure*}
\includegraphics[width=160mm]{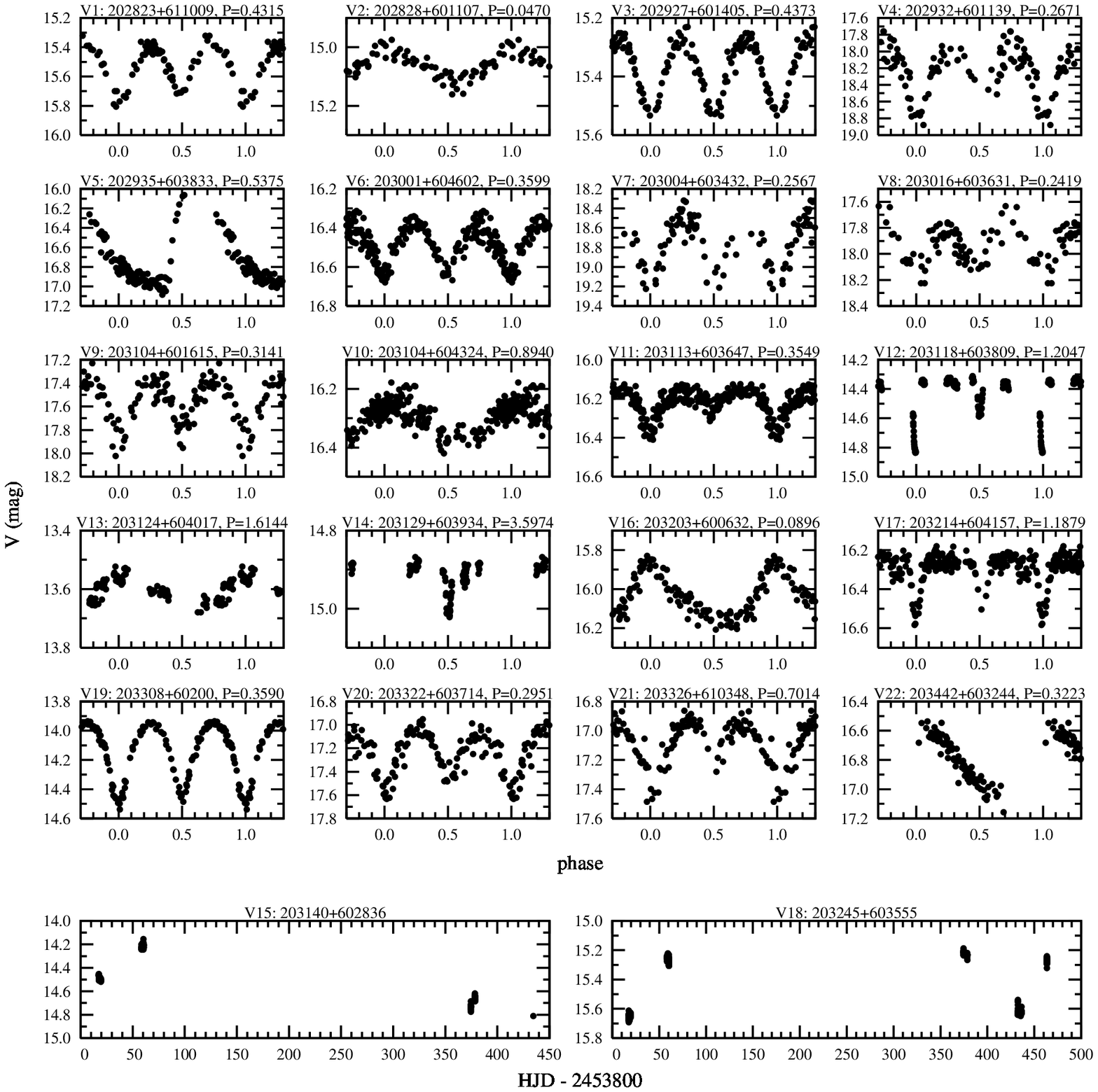}
\caption{The $V$ light curves of variables detected in the field of NGC~6939.}
\label{fig1}
\end{figure*}

\begin{figure}
\includegraphics[width=80mm]{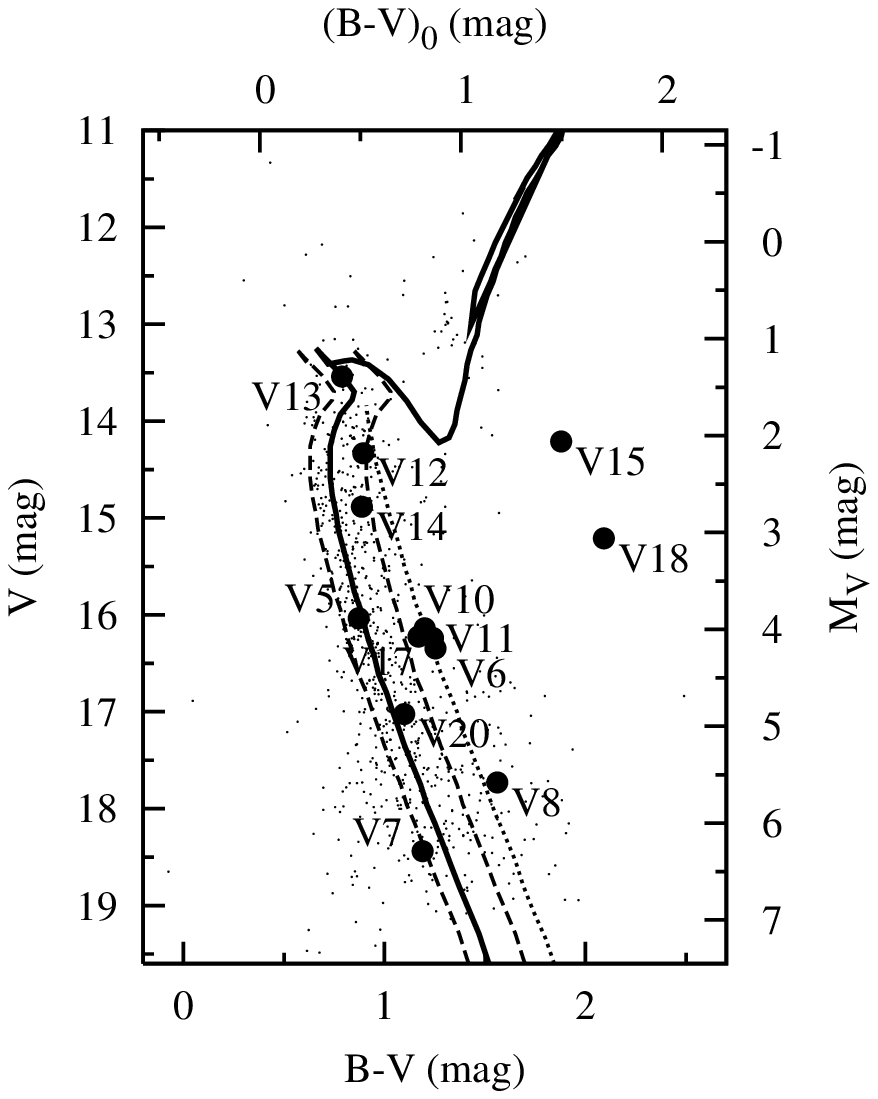}
\caption{Colour-magnitude diagram for NGC~6939 with marked individual variables located within cluster limiting radius. The continuous line denotes the isochrone taken from Maciejewski \& Niedzielski (2007). Top and right axes are rescaled to dereddened color index $(B-V)_{0}$ and absolute magnitudes $M_{\mathrm{V}}$, respectively.}
\label{fig2}
\end{figure}

As a result of our survey 22 variable stars in the field of NGC~6939 were detected. Four of them (V11--14) were discovered by Robb \& Cardinal (1998). All variables are listed in Table~\ref{table1} and their light curves in V band are displayed in Fig.~\ref{fig1}. Thirteen of them are located within the distance smaller than the limiting radius $r_{\mathrm{lim}}=15.2'$ from the cluster center. To verify the possibility that they belong to the cluster  $(B-V) \times V$ colour-magnitude diagram (CMD) was plotted in Fig.~\ref{fig2}, based on data taken from the OCS,  for stars within the radius of 8' from the cluster center -- a value smaller than cluster limiting radius to minimize the background star contamination. The isochrone reflecting cluster parameters given in Maciejewski \& Niedzielski (2007) is drawn with a continuous line. The errors of the isochrone fit in $E(B-V)$, that include also reported variable reddening in the cluster region, are marked by two dashed lines. 

As one can note, most of stars are situated between these two lines. The dotted line emblems the border of possible binary systems. The  variables which localization on the sky suggests cluster membership are marked with filled circles. 
Two of them V15 and V18 (both red variables of SR or irregular type) can be immediately deleted from the list of possible cluster members. The proximity to the cluster CMD of the remaining 11 stars allowed us to discuss their membership probability to NGC~6939.

\subsection{Individual variable stars}
V5 is classified as an RRAB pulsating star. It is located in the far outskirts of the cluster but very close to the main sequence in the CMD. We conclude however,  that V5 is not a cluster member  because RR Lyrae stars are not expected to be present in relatively young open clusters because they are pulsating horizontal branch stars with a mass of around 0.5 $M_{\odot}$.
  
V6 is an eclipsing system of EW type, located almost 13' from the cluster center. In the CMD the system is situated at a significant distance from the main sequence, just beyond the sketched border. That makes its cluster  membership unlikely. 

V7 is a contact binary system belonging rather to the Galactic background than to the cluster. The system is distant from the cluster center and located below the main sequence in the CMD, just behind sketched border. 

V8 is another eclipsing EW system which probability of cluster membership was estimated to unlikely low due to its location outside clusters main sequence estimated errors. 

The variable V10 reveals low amplitude brightness changes that allowed us to classify it as a pulsating star of RR Lyrae type. That suggests that the star is not a member of the cluster.

The variable V11 was detected by Robb \& Cardinal (1998) where it was designated as s154 and classified as a contact binary. Our data clearly indicates that V11 is an eclipsing binary of EB type. It is located near the cluster center. However, as one can see in Fig.~\ref{fig2}, its location in the CMD -- just beyond the sketched border -- makes its membership unlike.

The light variations of V12 were discovered by Robb \& Cardinal (1998) who classified it as an Algol-type binary system with a period of $4.954\pm0.003$ days. In our data only two minima are recorded what made our period determination uncertain. The system is located near the cluster center in the sky, as well as situated within acceptable distance from the main sequence in the CMD. Therefore we conclude that V12 is likely to be a member of NGC~6939.  

The variability of V13 was found by Robb \& Cardinal (1998). These authors designated it as s28, determined period of variation of $1.30\pm0.02$ day and stated that the star lied in the red giant part of the HR diagram. However, our photometry indicates that V13 is located near the turn-off point in the CMD. Variations of brightness indicate that the variable is a short-period pulsating star with phased light-curve morphology typical for cepheids. V13 is therefore probably a Galactic-background variable star because variables of that type are expected in clusters younger than NCG~6939.

Variability of V14 was detected by Robb \& Cardinal (1998) who recorded it as s98. These authors classified the variable as an Algol-type eclipsing system with a period of $3.598\pm0.003$ days. The primary minimum is about 0.35 mag deep while the secondary  about 0.1 mag. During our campaign only secondary minimum was observed twice. The system is located near the cluster center, as well as near the cluster main sequence. That makes the membership  of V14 to NGC~6939 likely.

V17 is a one more Algol-type eclipsing system located near the cluster center. Its location in the CMD, well within estimated errors, makes the star cluster membership likely. 

The contact system V20 is located in the outskirts of the cluster. However, its position in the CMD suggests that the star is a cluster member. As it is shown in Fig.~\ref{fig2}, the star is located in the vicinity of the main sequence, just above it. Thus we conclude that V20 is likely to  belong to NGC~6939.   

Variable s11, reported already in Robb \& Cardinal (1998), was not monitored in our survey due to saturation. Variability of s33 of Robb \& Cardinal (1998) was not confirmed. The brightness of the star appeared to be constant.

\subsection{Eclipsing systems}

\begin{table}
\centering
\caption{Absolute magnitudes of EW and EB systems located within cluster radius calculated in two different ways (see text for details).}
\label{table2}
\begin{tabular}{lcc}
\hline
ID &  $M_{\mathrm{V}}^{\mathrm{iso}}$ (mag) & $M_{\mathrm{V}}^{\mathrm{EW}}$ (mag) \\
\hline
V6  & 4.20 & 4.72 \\
V7  & 6.29 & 5.19 \\
V8  & 5.58 & 6.42 \\
V11 & 3.99 & 4.59 \\
V20 & 4.87 & 4.64 \\
\hline
\end{tabular}
\end{table}

To verify membership of contact binaries and short-period EB eclipsing systems their absolute magnitudes were calculated in two ways under an assumption that variables are cluster's members. The absolute magnitude $M_{\mathrm{V}}^{\mathrm{iso}}$ of the system may be estimated  basing on its maximum brightness $V_{\mathrm{max}}$ and cluster's distance modulus $(m-M)$ taken from Maciejewski \& Niedzielski (2007) as
\begin{equation}
 \label{m_izo}
    M_{\mathrm{V}}^{\mathrm{iso}}=V_{\mathrm{max}} - (m-M) \, . \;
\end{equation}
On the other hand, the absolute magnitude $M_{\mathrm{V}}^{\mathrm{EW}}$ can also be obtained from the empirical formula 
\begin{equation}
 \label{m_ruc}
    M_{\mathrm{V}}^{\mathrm{EW}}=-4.44 \log P + 3.02(B-V)_{0} + 0.12 \, , \;
\end{equation}
where $P$ is period of variation in days and $(B-V)_{0}$ is dereddened color index (Ruci\'nski i Duerbeck 1997, Ruci\'nski 2004). If a system belongs to the cluster both values would be identical within  typical error of 0.25 mag (Ruci\'nski 2004). The results of calculations are collected in Table~\ref{table2}. It is clear  that V20 is the only contact system that belongs to the cluster what support conclusions reached in discussion for individual variables.

\section{Summary and conclusions}

\begin{figure}
\includegraphics[width=65mm]{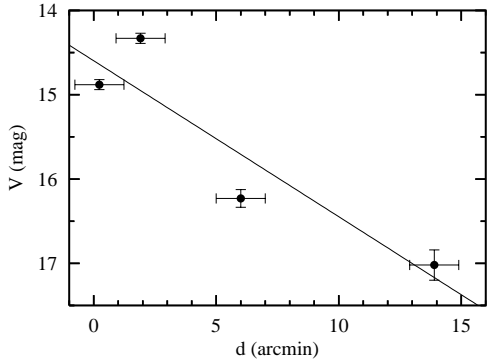}
\caption{Relation between maximum brightness of the cluster's eclipsing systems and their distance from the cluster center. A distance error of 1' -- the precision of cluster's center redetermination (Maciejewski \& Niedzielski 2007) was adopted. The value of $3 \sigma$ of typical accuracy obtained in our survey in V band was taken as a magnitude error.}
\label{fig3}
\end{figure}

  The results of a CCD photometric survey of the field around NGC~6939 were presented. Twenty two variable stars were detected, only four of them were previously known. Only 4 eclipsing systems (3 detached and 1 contact binary) from 13 variables located within a cluster radius occurred to be likely cluster members. Analysis of absolute magnitude of the newly detected contact system gives the distance to the cluster $D = 1.74 \pm 0.20$ kpc which value is in agreement with previous studies. 
  
The population of contact binary systems was found to be small in NGC~6939. This observation is in agreement with results obtained for other open cluster of similar age (e.g. Mazur at al. 1995 and references therein). Rich populations of contact binaries occur in old clusters of age of a few Gyr. The relatively rich population of short-period ($P<5$ days) detached systems was found. One cannot exclude that these systems are progenitors of W UMa stars which are formed as a result of angular momentum loss in initially detached systems (Vilhu 1982).  

It is worth noting that the relation between maximum brightness of the cluster's eclipsing systems (V12, V14, V17, and V20) and their distance from the cluster center was noticed (Fig.~\ref{fig3}). The relation $V = (0.19 \pm 0.06) d + (14.60 \pm 0.44)$ with the correlation coefficient of 0.92 was obtained as a result of the linear least-square best fit. Maximum brightness of eclipsing systems can be interpreted as a rough approximation of system's total mass in a star cluster. As a result of the cluster internal dynamics more massive (i.e. brighter) systems should occupy the central part of the cluster while the less massive (i.e. fainter) ones -- the outer region. The relation presented in Fig.~\ref{fig3} seems to confirm mass segregation among cluster's binary systems.

\textit{Acknowledgements}:
We thank Svetlana Boeva for the help in a part of the observations. This paper is a result of PAN/BAN exchange and joint research project \textit{Spectral and photometric studies of variable stars}. This research is also supported by UMK grant 369-A and Polish Ministry of Science and Higher Education grant 1P02D00730.

\end{document}